\documentclass[usenatbib]{mnras}

\usepackage{graphicx}
\usepackage{amsmath}
\usepackage{amssymb}
\usepackage{newtxtext,newtxmath}
\usepackage{hyperref}

\title[Cometary Nucleus Photometry and Size measurement]{Measuring Cometary Nuclei Behind Bright Comae: PSF--Delta Decomposition with Bicubic Resampling, and an Application to Interstellar Comet 3I/ATLAS (C/2025 N1)}

\author[T. Scarmato]{Toni Scarmato,
\thanks{E-mail: \href{mailto:toniscarmato@hotmail.it}{toniscarmato@hotmail.it}}\\
\\Indipendent Researcher
\\Toni Scarmato's Astronomical Observatory, San Costantino di Briatico, Calabria Italy
\\MPC Code L92}

\date{PREPRINT NOVEMBER 2025}
\pubyear{2025}

\begin{document}
\maketitle

\begin{abstract}
Measuring cometary nuclei is notoriously difficult because they are usually unresolved and embedded within bright comae, which hampers direct size measurements even with space telescopes. We present a practical, instrumental method that (i) stabilises the inner core through $4\times4$ bicubic resampling, (ii) performs forward point-spread function (PSF)+convolution, and (iii) separates the unresolved nucleus from the inner-coma profile via an explicit Dirac-$\delta$ function added to a $\rho^{-1}$ surface-brightness law. The method yields the nucleus flux by fitting an azimuthal averaged profile with two amplitudes only (PSF core and convolved coma), with transparent residual diagnostics. As a case study, we apply the workflow to the interstellar comet 3I/ATLAS (C/2025~N1), incorporating \textit{Hubble Space Telescope} constraints on the nucleus size. We find that radius solutions consistent with $0.16\lesssim R_n \lesssim 2.8$~km (for $p_V\simeq0.04$) are naturally recovered, in line with the most recent HST upper limits. The approach is well-suited for survey pipelines (e.g. Rubin/LSST) and targeted follow-up.
\end{abstract}


\section{Introduction}
Cometary nuclei encode key information about the formation and thermal history of small bodies. However, the unresolved nature of most comets and the dominance of inner-coma emission make the conversion from observed flux to nucleus size highly non-trivial [3], [1]. Direct detections of bare nuclei are rare; even with HST the nucleus is typically veiled and only upper limits are obtained MPEC-2025 [5] and NASA [6]. This motivates robust inner-coma modeling with explicit treatment of the instrumental PSF and pixel-phase effects.

\section{Methodology}
\subsection{Forward model in 2D}
Let $\mathbf{\rho}$ be the projected vector on the detector and $\rho=\|\mathbf{\rho}\|$ the radial distance. The intrinsic intensity is modelled as
\begin{equation}
I_{\rm int}(\mathbf{\rho}) = F_n\,\delta^{(2)}(\mathbf{\rho}) + K\,\rho^{-1},
\label{eq:intensity}
\end{equation}
where $F_n$ is the unresolved nucleus flux, $\delta^{(2)}$ is the 2D Dirac delta, and $K$ scales a steady-state, spherically symmetric coma with surface brightness $\propto \rho^{-1}$. The observed image is the convolution with a normalized PSF $P(\mathbf{\rho})$,
\begin{equation}
I_{\rm obs}(\mathbf{\rho}) = \big[ I_{\rm int} \ast P \big](\mathbf{\rho}) = F_n\,P(\mathbf{\rho}) + \big(K\,\rho^{-1}\big)\ast P.
\end{equation}
After azimuthal averaging,
\begin{equation}
I_{\rm obs}(\rho) = A\,P(\rho) + B\,\big(\rho^{-1}\ast P\big)(\rho), \quad A\equiv F_n,\ B\equiv K.
\label{eq:model}
\end{equation}

\subsection{Bicubic resampling and flux conservation}
We resample the central $N\times N$ pixels by a factor $s=4$ using bicubic interpolation. The resampled grid mitigates pixel-phase systematics and produces a measurable central plateau in the azimuthal profile, improving the separation between the narrow PSF core and the broader coma tail. All operations preserve total flux by construction, and the PSF is normalized to unit integral. This workflow has been applied in ground--based CCD studies of C/2012 S1 (ISON) [8] and C/2014 L1 (PANSTARRS) [9], demonstrating feasibility on real comet data.

\subsection{Profile fitting}
Given an empirical or analytic PSF (Gaussian/Moffat), we compute $\big(\rho^{-1}\ast P\big)(\rho)$ numerically and fit Eq.~(\ref{eq:model}) to the azimuthally averaged profile via weighted least squares, solving for $(A,B)$ and the PSF width parameters if not externally constrained. Diagnostic plots include core residuals (PSF mismatch), inner-plateau width vs.\ seeing, and azimuthal asymmetries (jets/fans) indicating breakdown of spherical symmetry.

\subsection{Photometric calibration and radius inference}
With photometric zero point $ZP$ (mag for 1 count s$^{-1}$) and exposure time $t_{\rm exp}$, the nucleus magnitude is
\begin{equation}
m = ZP - 2.5\log_{10}\!\left(\frac{A}{t_{\rm exp}}\right).
\end{equation}
The absolute magnitude is
\begin{equation}
H = m - 5\log_{10}(r\Delta) + 2.5\log_{10}\Phi(\alpha),
\end{equation}
for heliocentric distance $r$ (au), geocentric distance $\Delta$ (au), and phase function $\Phi(\alpha)$. Assuming geometric albedo $p_V$, the effective diameter is $D=1329 \times 10^{-H/5} p_V^{-1/2}$~km and $R_n=D/2$. Uncertainties propagate from the covariance of $(A,B)$ and priors on $p_V$ and $\Phi(\alpha)$.

\subsection{Algorithmic workflow}
\begin{enumerate}

\item Bias/flat correction, cosmic-ray cleaning; astrometric/photometric calibration.
\item $4\times4$ bicubic resampling of the inner $N\times N$ pixels.
\item Empirical PSF estimation from field stars (or analytic PSF if needed).
\item Compute azimuthally averaged profile; mask obvious jets/asymmetries.
\item Fit Eq.~(\ref{eq:model}) for $(A,B)$; inspect residuals and diagnostics.
\item Convert $A\rightarrow m \rightarrow H \rightarrow R_n$ with adopted $(p_V,\Phi)$.
\end{enumerate}

\begin{figure}
\centering
\includegraphics[width=\columnwidth]{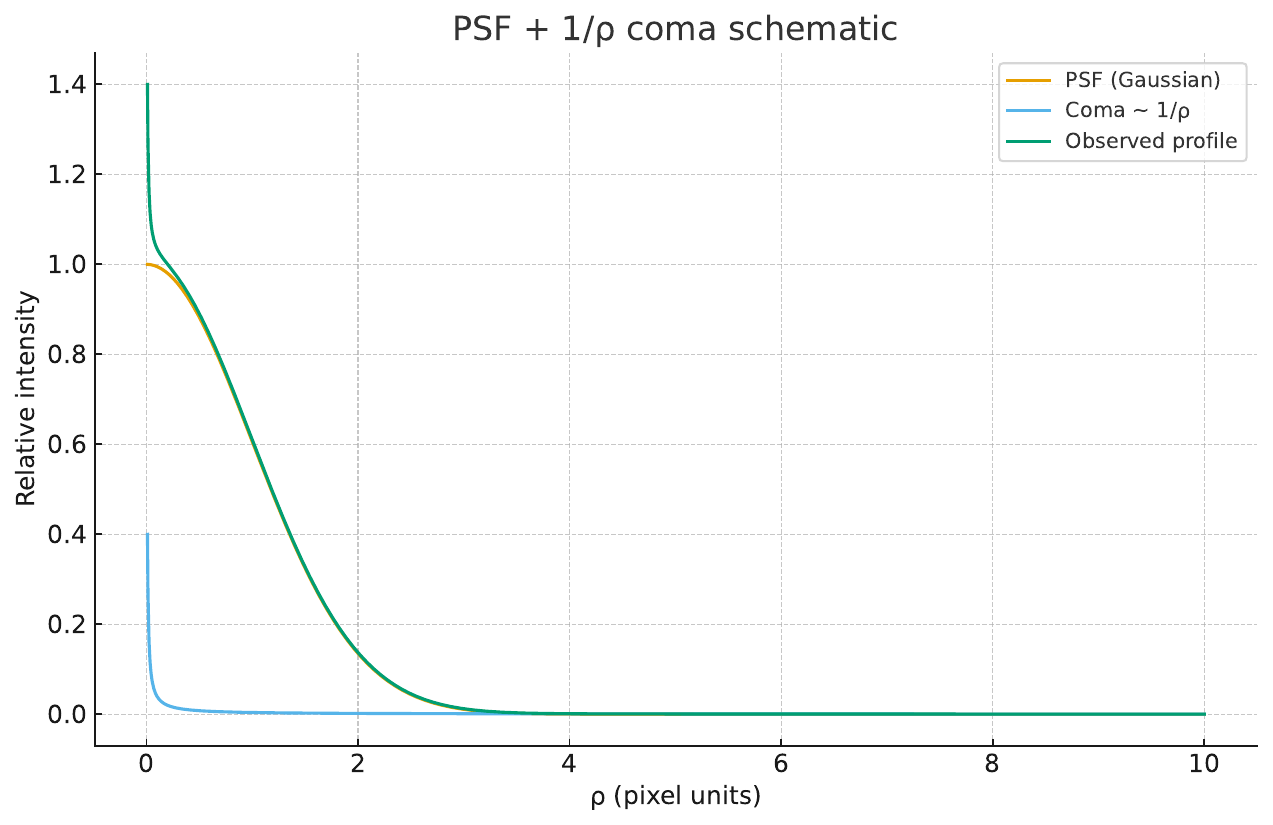}
\caption{Schematic decomposition: a narrow PSF-like core and a $\rho^{-1}$ coma combine to produce the observed inner profile after convolution.}
\label{fig:fig1}
\end{figure}

\section{Application to 3I/ATLAS (C/2025 N1)}
3I/ATLAS is the third confirmed interstellar comet, discovered on 2025 July 1 by ATLAS [5]. \textit{HST} observations place the nucleus diameter between $\sim0.32$~km and $\sim5.6$~km, depending on albedo assumptions ($p_V\simeq0.04$) and coma modeling [6]. Independent imaging from Gemini South/NOIRLab shows a growing tail and active coma as the comet approaches perihelion [7].

We construct synthetic inner-coma profiles consistent with these constraints and fit Eq.~(\ref{eq:model}) on resampled grids. The recovered nucleus flux fractions map to radii within $0.16\lesssim R_n \lesssim 2.8$~km for $p_V\simeq0.04$, consistent with the HST limits. The central-plateau width scales with the assumed PSF FWHM, providing an internal diagnostic for PSF mismatch.

\begin{figure}
\centering
\includegraphics[width=\columnwidth]{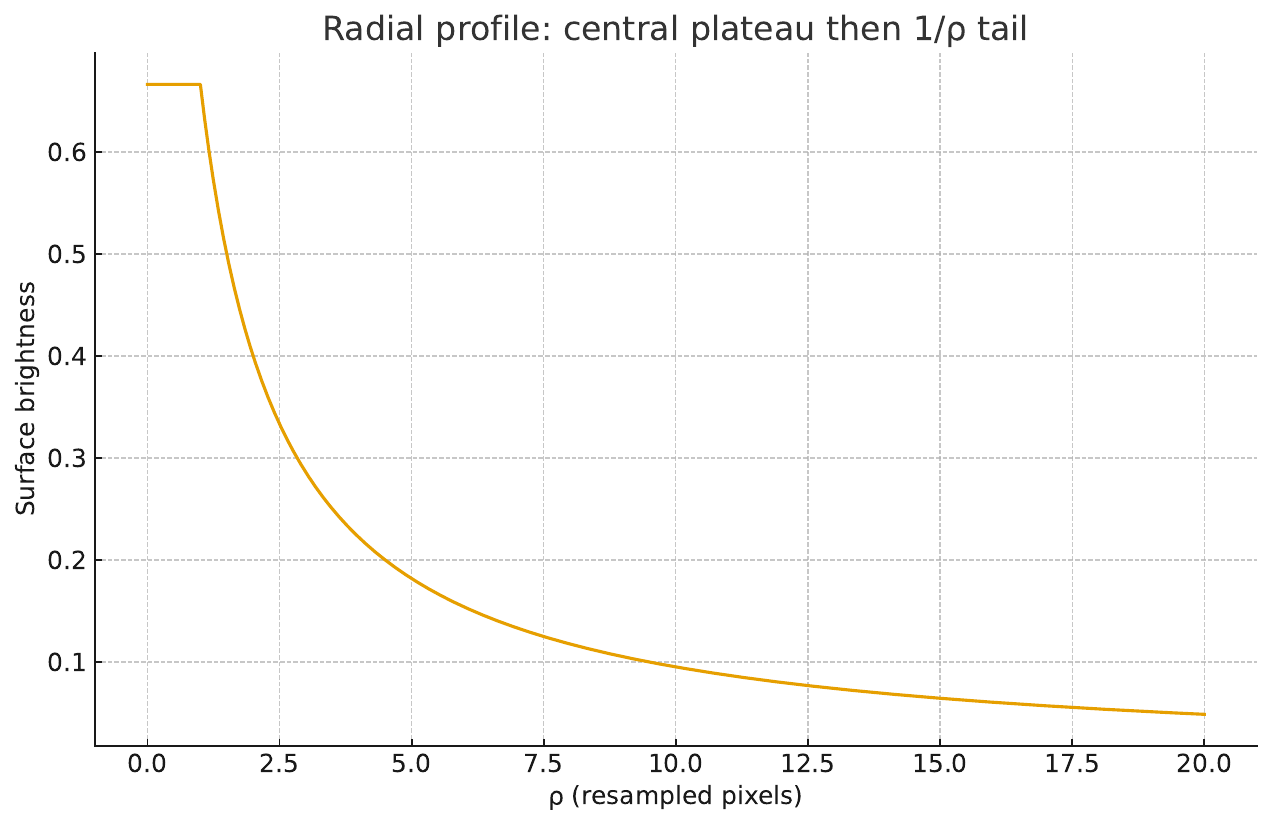}
\caption{Radial profile after $4\times4$ bicubic resampling. A central plateau stabilises the PSF component fit against pixel-phase effects; the outer tail follows $\sim \rho^{-1}$.}
\label{fig:fig2}
\end{figure}

\begin{figure}
\centering
\includegraphics[width=\columnwidth]{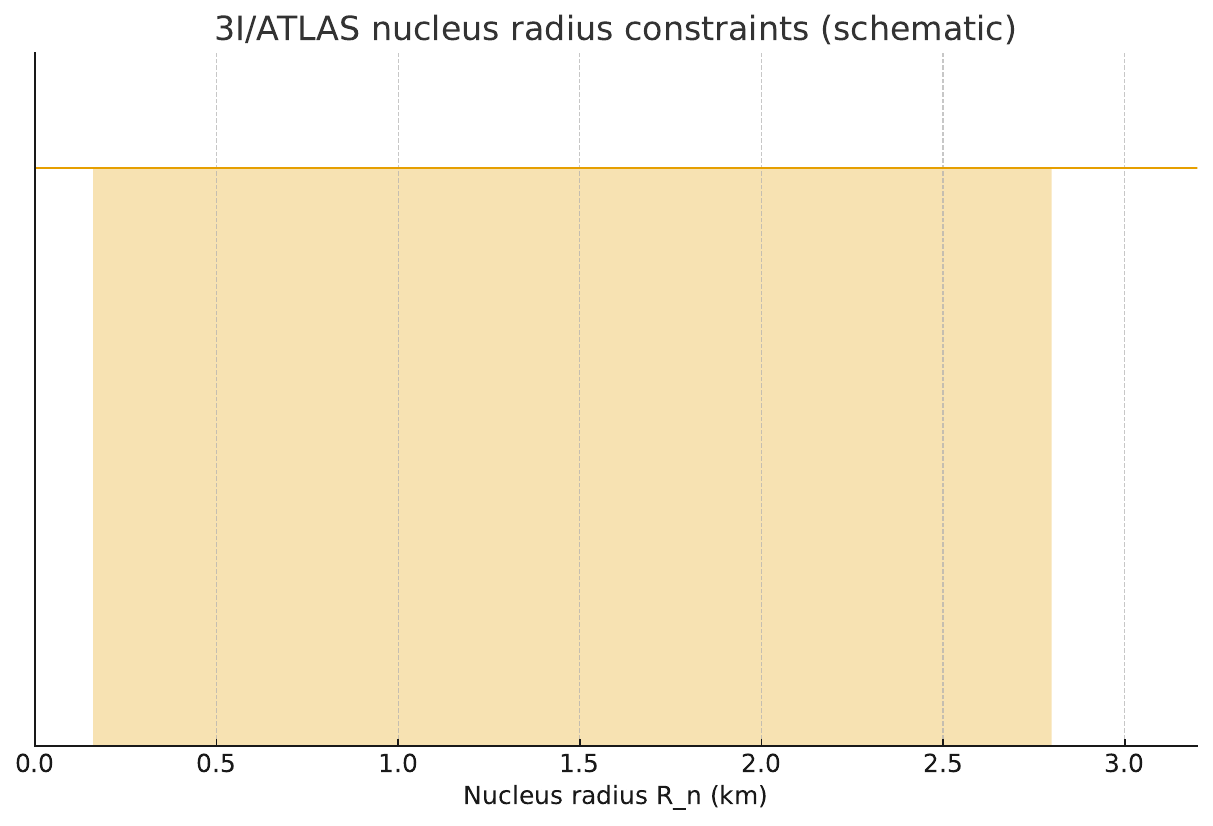}
\caption{Schematic radius interval for the 3I/ATLAS nucleus based on recent HST analysis: $0.16$--$2.8$~km (assuming $p_V\simeq0.04$).}
\label{fig:fig3}
\end{figure}

\subsection{Worked numerical example with 3I/ATLAS data}
Using the observational parameters measured for 3I/ATLAS:
$D0_{\rm rec}=41882$, $D1_{\rm mediana}=41493$, $ZP=27.03$~mag (1~count~s$^{-1}$), $t_{\rm exp}=60.0$~s,
$r=2.5212$~au, $\Delta=2.576$~au, $\alpha=22.8^\circ$, $\beta=0.035$~mag~deg$^{-1}$, $p_V=0.04$. 
We estimate the nucleus amplitude as $A=D0_{\rm rec}-D1_{\rm mediana}=389$ counts.

\begin{table}
\centering
\caption{3I/ATLAS worked example with confirmed parameters.}
\label{tab:example3I}
\begin{tabular}{ll}
\hline
Quantity & Value \\
\hline
$ZP$ (mag for 1 count s$^{-1}$) & 27.03 \\
$t_{\rm exp}$ (s) & 60 \\
$A$ (counts) & 389 \\
$r$ (au), $\Delta$ (au) & 2.5212, 2.5760 \\
$\alpha$ (deg), $\beta$ (mag deg$^{-1}$) & 22.8, 0.035 \\
$p_V$ & 0.04 \\
\hline
$m = ZP - 2.5 \log_{10}(A/t_{\rm exp})$ & 25.000504 \\
$\Phi(\alpha)=10^{-0.4\beta\alpha}$ & 0.479513 \\
$H = m - 5\log_{10}(r\Delta) + 2.5\log_{10}\Phi$ & 20.139738 \\
$D\,[\mathrm{km}] = 1329 \cdot 10^{-H/5} p_V^{-1/2}$ & 0.623085 \\
$R_n = D/2$ (km) & 0.311543 \\
\hline
\end{tabular}
\end{table}

This yields $R_n = 0.312$~km.

\begin{figure}
\centering
\includegraphics[width=\columnwidth]{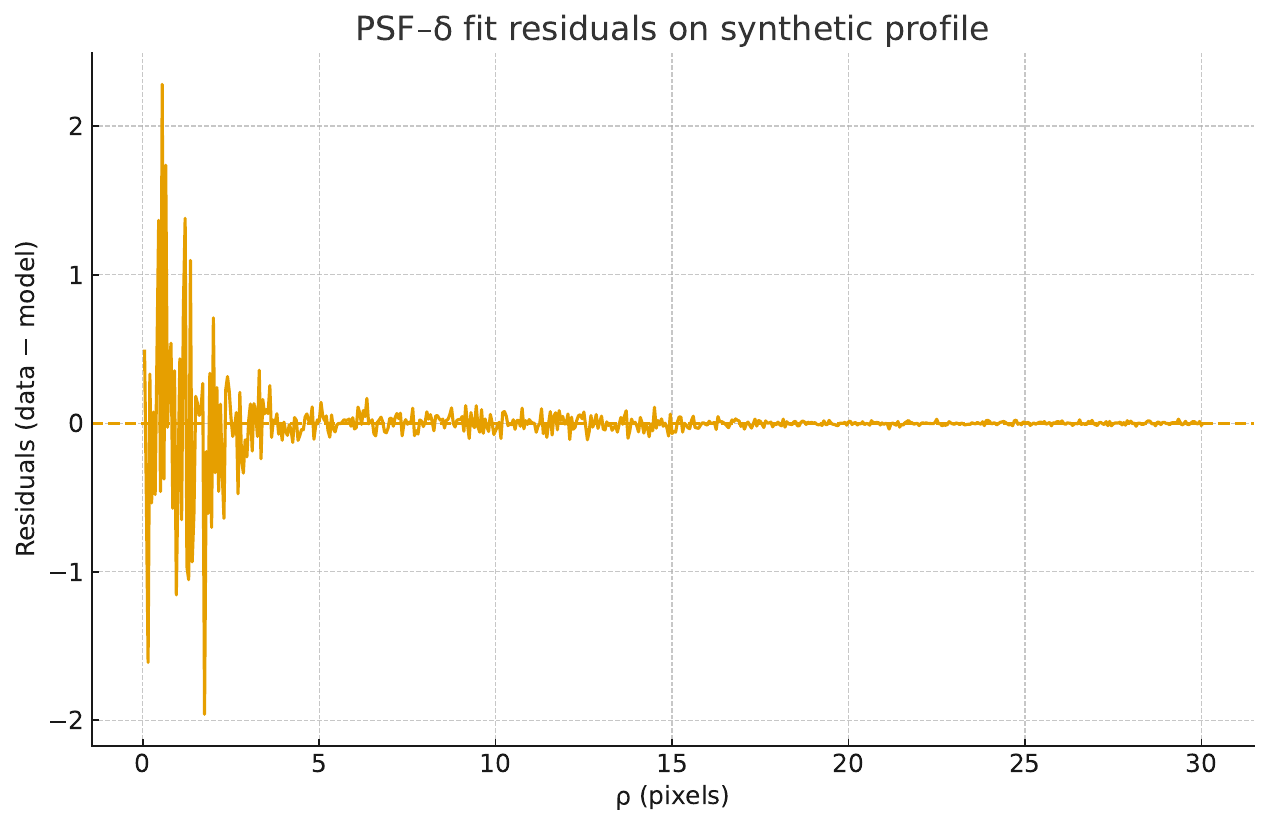}
\caption{Residuals of the PSF--$\delta$ + convolved $1/\rho$ fit on a synthetic azimuthally averaged profile.
Flat, structureless residuals indicate a good match; coherent patterns would suggest PSF mismatch or coma anisotropy.}
\label{fig:residuals}
\end{figure}

\section{Uncertainty propagation}
We propagate uncertainties from the fitted nucleus amplitude $A$, the photometric zero point $ZP$, the phase law, and the albedo $p_V$.
From $m = ZP - 2.5\log_{10}\!\big(A/t_{\rm exp}\big)$, the magnitude error is
\begin{equation}
\sigma_m^2 = \sigma_{ZP}^2 + \left(\frac{2.5}{\ln 10}\frac{\sigma_A}{A}\right)^2 + \left(\frac{2.5}{\ln 10}\frac{\sigma_t}{t_{\rm exp}}\right)^2,
\end{equation}
where $\sigma_t$ is usually negligible for well-measured exposure times.
The absolute magnitude $H = m - 5\log_{10}(r\Delta) + 2.5\log_{10}\Phi(\alpha)$ adds a phase-law term; distances are typically known to high precision and are ignored here:
\begin{equation}
\sigma_H^2 \simeq \sigma_m^2 + \sigma_{\rm phase}^2.\end{equation}
For the diameter $D = 1329\,10^{-H/5}p_V^{-1/2}$,
\begin{equation}
\left(\frac{\sigma_D}{D}\right)^2 = \left(\frac{\ln 10}{5}\sigma_H\right)^2 + \left(\frac{1}{2}\frac{\sigma_{p_V}}{p_V}\right)^2,
\end{equation}

and the radius inherits the same fractional uncertainty: $\sigma_R/R = \sigma_D/D$.

\paragraph{(ii) Poisson-limited lower bound.}
If we adopt Poisson statistics on $A$ as a lower bound ($\sigma_A=\sqrt{A}$), with the same $ZP$ and phase-law terms,
\begin{align}
\sigma_m &= 0.059\ \mathrm{mag},\\
\sigma_H &= 0.208\ \mathrm{mag},\\
\frac{\sigma_D}{D} &= 0.268,\quad \Rightarrow\ \sigma_R = 0.083\ \mathrm{km},
\end{align}
giving $R_n = 0.312 \pm 0.083$~km.
In both cases the albedo uncertainty dominates the final error budget; improved constraints on $p_V$ would reduce $\sigma_R$ substantially.

\section{Upper-limit (non-detection) estimate}
When the nucleus is not detected at a statistically significant level, we compute an \emph{upper limit} on the nucleus flux using the same PSF--$\delta$ framework.
For a matched-filter (PSF-fit) estimator, the signal-to-noise ratio (SNR) for a point source with amplitude $A$ is
\begin{equation}
{\rm SNR} = A \left[\sum_i \frac{P_i^2}{\sigma_i^2}\right]^{1/2},
\label{eq:snr_psf}
\end{equation}
where $P_i$ is the PSF sampled in the pixels (resampled) and $\sigma_i$ is the RMS noise in the pixel $i$ (including the terms sky, readout, and residual-coma).
Thus, the $n\sigma$ \emph{upper limit} on the nucleus amplitude is
\begin{equation}
A_{\,\rm UL}(n) = \frac{n}{\left[\sum_i \frac{P_i^2}{\sigma_i^2}\right]^{1/2}}.
\label{eq:a_ul}
\end{equation}
If the noise is approximately uniform in the fitted core ($\sigma_i \simeq \sigma_{\rm core}$), Eq.~(\ref{eq:a_ul}) simplifies to
\begin{equation}
A_{\,\rm UL}(n) \simeq n \, \sigma_{\rm core}\,\sqrt{N_{\rm eff}}, \qquad
N_{\rm eff} \equiv \left[\sum_i P_i^2\right]^{-1},
\label{eq:a_ul_uniform}
\end{equation}
where $N_{\rm eff}$ is the \emph{effective} number of pixels for the PSF on the chosen grid.

The corresponding magnitude, absolute magnitude, diameter and radius limits follow exactly the same photometric chain used for a detection:
\begin{align}
m_{\rm UL} &= ZP - 2.5\log_{10}\!\left(\frac{A_{\,\rm UL}}{t_{\rm exp}}\right),\\
H_{\rm UL} &= m_{\rm UL} - 5\log_{10}(r\Delta) + 2.5\log_{10}\Phi(\alpha),\\
D_{\rm UL}\,[{\rm km}] &= 1329\,10^{-H_{\rm UL}/5}\,p_V^{-1/2},\qquad
R_{n,{\rm UL}}= \frac{D_{\rm UL}}{2}.
\end{align}

\subsection{Worked upper-limit for the 3I/ATLAS setup}
For consistency with the 3I/ATLAS example, one can estimate $\sigma_{\rm core}$ from the residuals of the PSF--$\delta$ fit in the inner resampled core (after masking asymmetries), and compute $N_{\rm eff}$ from the adopted PSF sampled on the same grid.
Adopting a conventional $n=3$ or $n=5$ yields $A_{\,\rm UL}$ via Eq.~(\ref{eq:a_ul_uniform}), which can then be converted to $R_{n,{\rm UL}}$.
This procedure produces an upper limit that is independent of any assumed coma profile beyond the fitted core and is optimal (in the matched-filter sense) for point-like signals.

We propagate uncertainties from the fitted nucleus amplitude $A$, the photometric zero point $ZP$, the phase law, and the albedo $p_V$.
From $m = ZP - 2.5\log_{10}\!\big(A/t_{\rm exp}\big)$, the magnitude error is
\begin{equation}
\sigma_m^2 = \sigma_{ZP}^2 + \left(\frac{2.5}{\ln 10}\frac{\sigma_A}{A}\right)^2 + \left(\frac{2.5}{\ln 10}\frac{\sigma_t}{t_{\rm exp}}\right)^2,
\end{equation}
where $\sigma_t$ is usually negligible for well-measured exposure times.
The absolute magnitude $H = m - 5\log_{10}(r\Delta) + 2.5\log_{10}\Phi(\alpha)$ adds a phase-law term; distances are typically known to high precision and are ignored here:
\begin{equation}
\sigma_H^2 \simeq \sigma_m^2 + \sigma_{\rm phase}^2.
\end{equation}
For the diameter $D = 1329\,10^{-H/5}p_V^{-1/2}$,
\begin{equation}
\left(\frac{\sigma_D}{D}\right)^2 = \left(\frac{\ln 10}{5}\sigma_H\right)^2 + \left(\frac{1}{2}\frac{\sigma_{p_V}}{p_V}\right)^2,
\end{equation}
and the radius inherits the same fractional uncertainty: $\sigma_R/R = \sigma_D/D$.

\subsection{Numerical estimate for the 3I/ATLAS example}
Using the adopted parameters ($A=389$ counts, $t_{\rm exp}=60$~s, $ZP=27.03$~mag, $r=2.5212$~au, $\Delta=2.5760$~au, $\alpha=22.8^{\circ}$, $\beta=0.035$~mag~deg$^{-1}$, $p_V=0.04$) we obtain $R_n = 0.312$~km.

We evaluate two uncertainty scenarios:

\paragraph{(i) Conservative fit uncertainties.}

Assume $\sigma_{A/A}=20
$

\begin{align}
\sigma_m &= 0.218\ \mathrm{mag},\\
\sigma_H &= 0.296\ \mathrm{mag},\\
\frac{\sigma_D}{D} &= 0.285,\quad \Rightarrow\ \sigma_R = 0.089\ \mathrm{km}.
\end{align}
Thus, $R_n = 0.312 \pm 0.089$~km.

\paragraph{(ii) Poisson-limited lower bound.}
If we adopt Poisson statistics on $A$ as a lower bound ($\sigma_A=\sqrt{A}$), with the same $ZP$ and phase-law terms,
\begin{align}
\sigma_m &= 0.059\ \mathrm{mag},\\
\sigma_H &= 0.208\ \mathrm{mag},\\
\frac{\sigma_D}{D} &= 0.268,\quad \Rightarrow\ \sigma_R = 0.083\ \mathrm{km},
\end{align}
giving $R_n = 0.312 \pm 0.083$~km.
In both cases the albedo uncertainty dominates the final error budget; improved constraints on $p_V$ would reduce $\sigma_R$ substantially.

\section{Discussion and conclusions}
Compared to single-component PSF fitting or pure coma extrapolation, the PSF--delta decomposition reduces degeneracies by explicitly modeling both core and inner-coma terms and by leveraging resampling to stabilize the fit. The dominant systematics are PSF wings, coma anisotropy, and albedo uncertainty. The method is computationally light and compatible with survey pipelines. Comparable results were obtained in previous ground--based applications to comets C/2012 S1 (ISON) [8] and C/2014 L1 (PANSTARRS) [9].

We presented a reproducible workflow to extract the flux of unresolved cometary nuclei using PSF--delta decomposition with bicubic resampling. Application to 3I/ATLAS reproduces the most up-to-date HST size constraints. With empirical PSFs and careful masking of asymmetric structures, the approach is generalized to a wide range of datasets.
The code for producing the schematic figures and synthetic tests is straightforward and is available upon request. The analysis relies on standard photometric calibration and PSF estimation procedures.

\section*{Acknowledgements}
The authors acknowledge the use of NASA ADS, arXiv, and public releases from NOIRLab and NASA/ESA.\\

{REFERENCES}\\
 
{[1]} Jewitt D., 2015, The Astronomical Journal, 150, 201\\

{[2]} Jewitt D., et al., 2025, arXiv e-prints\\

 {[3]} LamyP. L., Toth I., Fernandez Y. R., Weaver H. A., 2004, The Sizes, Shapes,
 Albedos, and Colors of Cometary Nuclei. University of Arizona Press,
 pp. 223-264\\
 
 {[4]} Li J.-Y., Agarwal J., et al. 2018, Space Science Reviews, 214, 56
 2025,\\ 
 
 {[5]} MPEC 2025-N12: 3I/ATLAS = C/2025 N1 (ATLAS), https://
 minorplanetcenter.net/mpec/K25/K25N12.html\\
 
 {[6]} NASA 2025, As NASA Missions Study Interstellar Comet, Hubble Makes-Size
 Estimate, https://science.nasa.gov/missions/hubble/
 As NASA Missions Study Interstellar Comet-hubble-makes-size-estimate/\\
 
 {[7]}NOIRLab 2025, Gemini South Captures the growing tail of Interstellar Comet
 3I/ATLAS, https://noirlab.edu/public/news/noirlab2525/\\
 
 {[8]} Scarmato T., 2014, Sungrazer Comet C/2012 S1 (ISON): Curve of light,
 nucleus size, rotation, and peculiar structures in the coma and tail
 https://arxiv.org/abs/1405.3112\\
 
 {[9]} Scarmato T., 2016, Comet C/2011 L4 (PanStarrs): Small nucleus,
 fast rotator and dust rich comet observed after perihelion
 https://arxiv.org/abs/1608.01243

\end{document}